# Characterization and Application of Hard X-Ray Betatron Radiation Generated by Relativistic Electrons from a Laser-Wakefield Accelerator

Michael Schnell,[1,*] Alexander Sävert,[1] Ingo Uschmann,[1,2] Oliver Jansen,[3] Malte Christoph Kaluza,[1,2] and Christian Spielmann[1,2]

[1] Institute of Optics and Quantum Electronics, Abbe Center of Photonics, Friedrich Schiller University, Max-Wien Platz 1, 07743 Jena, Germany

[2] Helmholtz Institute Jena, Friedrich Schiller University, Fröbelstieg 3, 07743 Jena, Germany

[3] Institute for Laser- and Plasmaphysics, Heinrich-Heine University Düsseldorf, 40225 Düsseldorf, Germany

[*] Michael.Schnell.1@uni-jena.de

**Abstract**

**The necessity for compact table-top x-ray sources with higher brightness, shorter wavelength and shorter pulse duration has led to the development of complementary sources based on laser-plasma accelerators, in contrast to conventional accelerators. Relativistic interaction of short-pulse lasers with underdense plasmas results in acceleration of electrons and in consequence in the emission of spatially coherent radiation, which is known in the literature as betatron radiation. In this article we report on our recent results in the rapidly developing field of secondary x-ray radiation generated by high-energy electron pulses. The betatron radiation is characterized with a novel setup allowing to measure the energy, the spatial energy distribution in the far-field of the beam and the source size in a single laser shot. Furthermore, the polarization state is measured for each laser shot. In this way the emitted betatron x-rays can be used as a non-invasive diagnostic tool to retrieve very subtle information of the electron dynamics within the plasma wave. Parallel to the experimental work, 3D particle-in-cell simulations were**

**performed, proved to be in good agreement with the experimental results.**

1. Introduction

Laser-plasma based electron accelerators are described in detail in a number of review articles e.g. by *Esarey et al.*[1] and *Krushelnick et al.*[2]. In all of these publications the authors highlight the importance of the setup for realizing table-top sources of relativistic electron bunches for future real world applications. Along with the longitudinal acceleration of the electrons (within the plasma) to relativistic energies[3-8] the electrons also undergo a transverse oscillation due to transverse electro-magnetic fields associated with a radial charge separation in the plasma wave[9].

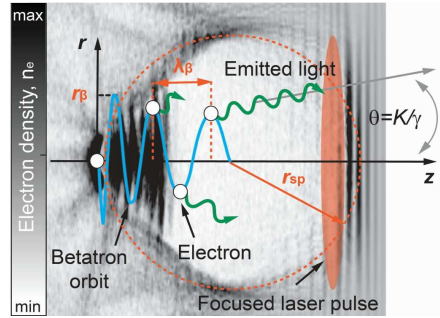

*Figure 1 (color online) Parameter for the generation of the betatron radiation from relativistic oscillating electrons. The laser propagates from left to right (red); electrons are expelled from the focal region by the laser pulse hence exciting a plasma wave in its wake. This plasma wave can break after strong excitation and electrons are injected into the associated electric field (the wakefield) and oscillate (blue trajectory) with the betatron amplitude, $r_\beta$, and wavelength, $\lambda_\beta$ (orange). The ion-cavity is described by a sphere of radius $r_{sp}$ (orange dashed circle). White circles show the radiation points where the electrons radiate light in the forward direction in a cone of divergence characterized by $\theta$ (dark green emitted light is shown in the lab frame).*

In the bubble regime[10], there exists a transverse electric field where the ion-cavity is described as a sphere of radius $r_{sp}$ (Fig. 1) with an electron density of $n_e$ (number of electrons per unit volume). For visualisation, in a cylindrical coordinate system r is the transverse distance from the laser propagation axis and r = 0 corresponds to the laser axis, where the ion-cavity is centred and the transverse field is zero (z-axis in Fig. 1). The advantage is that plasmas supports multimega-Gauss magnetic fields[11] resulting in a much shorter effective wiggler period on the order of a few hundred micrometer[12]. The basic idea for the generation of laser-plasma based betatron radiation was first reported in 2004 by *Kiselev et al.*[13] and experimentally demonstrated by *Rousse et al.*[14]. The radial space charge field will force the highly relativistic electrons to perform a transverse oscillation with the betatron wavelength[9,15-17] which leads to the emission of the betatron radiation, whose characteristics are similar to the synchrotron radiation in the wiggler regime[18] at the fundamental wavelength of $\lambda_0 \simeq 2.36 \times 10^{10} \sqrt{\frac{1}{\gamma^3 (n_e/cm^{-3})}} \mu m$, where $\gamma$ is the relativistic Lorentz factor. The ion-cavity shown in Figure 1 acts as a plasma undulator with a betatron strength parameter in practical units written as

$$K = \gamma\theta \simeq 1.33 \times 10^{-10}\sqrt{\gamma(n_e/cm^{-3})}(r_\beta/\mu m), \quad (1)$$

where $\theta$ is the half angle of divergence and $r_\beta$ the betatron amplitude of the electron trajectory. This oscillation of the relativistic electron is similar to that of an electron oscillating in a conventional undulator or in a wiggler. So it makes sense to use also here the characteristic parameters like electron period or the strength parameter, which are commonly used in the synchrotron and Free-electron laser ("FEL") community. Consequently, the radiation is emitted in forward direction in a narrow cone and can be characterized by the amplitude of the strength parameter, K (Eq. 1). In the bubble regime[19-21] K is typically larger than one. The continuous x-ray radiation centred on the observation direction **n** can be described by the radiated spectrum of a single electron on an arbitrary electron trajectory **r**(t)[22]

$$\frac{d^2I}{d\omega d\Omega} = \frac{e^2}{16\pi^3\varepsilon_0 c} \times \left| \int_{-\infty}^{+\infty} \exp(-i\omega[t - \vec{n}\cdot\vec{r}(t)/c]) \frac{\vec{n}\times\left[(\vec{n}-\vec{\beta})\times\dot{\vec{\beta}}\right]}{(1-\vec{\beta}\cdot\vec{n})^2} dt \right|^2. \quad (2)$$

Here, I is the radiated energy emitted into the solid angle $d\Omega$ within a spectral band $d\omega$ centred on the frequency $\omega$, e is the elementary charge, $\varepsilon_0$ the vacuum permittivity and c the speed of light in vacuum. The emitted betatron spectrum depends on (and can be controlled by) the electron velocity normalized to the speed of light in vacuum, $\vec{\beta}$, and the electron trajectory. For an asymptotic behaviour of the radiated spectrum observed on-axis θ=0, Equation 2 can be simplified to[23]

$$\left.\frac{d^2I}{d\omega d\Omega}\right|_{\theta=0} \simeq N_\beta \frac{3e^2}{2\pi^3\hbar\varepsilon_0 c}\gamma^2\xi^2\mathcal{K}_{2/3}^2(\xi), \qquad (3)$$

where $N_\beta$ is the number of oscillations, $\hbar$ the reduced Planck constant, $\mathcal{K}_{2/3}$ the modified Bessel function of the second kind and $\xi = E/E_{crit}$. Here $E_{crit}$ represents the energy, within the distribution, where half of the radiated power is below $E_{crit}$ and the other half lies above $E_{crit}$. This critical energy is defined in practical units by,

$$E_{crit} = \hbar\omega_{crit} \simeq 5\times 10^{-24}\gamma^2(n_e/cm^{-3})(r_\beta/\mu m)keV. \qquad (4)$$

The combination of the broad spectral range and inherent femtosecond-timescale synchronization of the electron and x-ray source with respect to the driving laser pulse makes this source suitable to monitor the motion of atoms and electrons on femtosecond-timescales within ultrafast, time-resolved, pump-probe experiments. Furthermore, electron and x-ray beams can also be used to investigate ultrafast laser-matter interactions with Ångstrom spatial resolution.

So far, betatron radiation has been successfully used for single-shot phase contrast imaging of biological samples within a compact setup[17] and for the application of x-ray absorption spectroscopy[24]. These sources are also of interest for high-resolution gamma radiography[25,26] in materials science and for femtosecond x-ray crystallography[27]. As presented in this article, the betatron radiation can also act as a non-invasive diagnostic to retrieve information on the acceleration dynamics of electron injection in laser-plasma accelerators[28-30]. Moreover, the measurement and the control of the betatron radiation's polarization state by steering the electrons inside the plasma will have a major impact on the ongoing efforts towards the realization of novel, laser-based particle accelerators.

The outline of the article will be as follows. After this short overview, a description of the experimental setup for the three-dimensional particle-in-cell code, called "Virtual Laser-Plasma Lab" (VLPL)[31], is given. This code helps to understand the physical mechanism involved in the laser-plasma interaction process and to confirm the experimental results presented in this article. The third section describes the entire experimental setup, followed by the evaluation of the obtained results including the betatron x-ray characterization and application as a diagnostic tool for electron injection and for the acceleration mechanism in laser-plasma accelerators together with results of the three-dimensional particle-in-cell code. After the conclusion we present in a short outlook the future experimental approaches and applications based on our findings.

## 2. Experimental setup for the PIC code

To simulate our experiments we have used the following geometry and settings which we have used in the experiments, too. A linearly polarized laser pulse with a normalized, relativistic amplitude, $a_{0,peak}$, was focused into an underdense plasma (values are given in the respective section, Fig. 2).

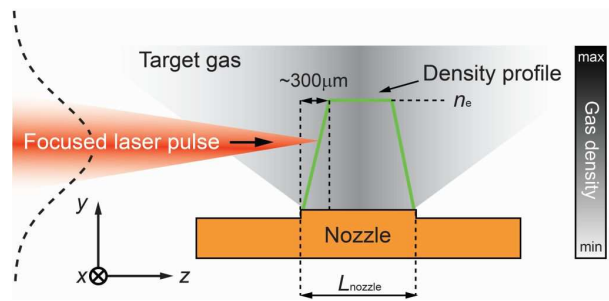

*Figure 2 (color online) Schematic of the simulation setup for the 3D-PIC code. The laser propagates from left to right and is focused into a gas density distribution, generated by a nozzle. The green solid line is the approximated electron density input profile for the PIC simulation.*

The plasma density increases linearly from zero to the nominal-value $n_e$ within the first 300μm, is then constant before it decreases to zero. Here, $n_e$ and $L_{nozzle}$ is the electron density and the nozzle exit diameter, respectively. The size of the simulation box is 40x$\lambda$ in all three dimensions (x, y and z). In the laser's propagation direction, the grid size was given by $\lambda/10$ (z-direction) and in the two transversal dimensions it was $\lambda/4$. The incident laser pulse had a Gaussian intensity distribution in the transverse direction (y-axis) and a cosine distribution in the propagation direction (z-axis). Other than relativistic self-focusing, the simulations also included the recoil force acting on an accelerating electron caused by the emitted radiation, also called "radiation reaction", during which the properties of the betatron radiation can be derived. One varying parameter was the peak amplitude of the normalized vector potential, $a_{0,peak}$, which is related to the laser's peak intensity, $I_{peak}$. The most important simulation parameters and numerical results are summarized in their respective sections.

## 3. Experimental setup and diagnostic

All experiments described in this article were carried out at the multi-TW Ti:sapphire chirped-pulse amplification[32] laser-system (called the "JETI") in Jena, Germany. This system delivered pulses with a duration of $\tau_{FWHM}$=30fs with an on-target energy of about 750mJ at a central wavelength of $\lambda$=800nm. The pulses with a beam diameter of roughly 50mm at full-width at half-maximum (FWHM) were focused by an F/13 off-axis parabolic mirror into the leading edge of a pulsed, super-sonic gas jet[33] of helium or hydrogen (Fig.3).

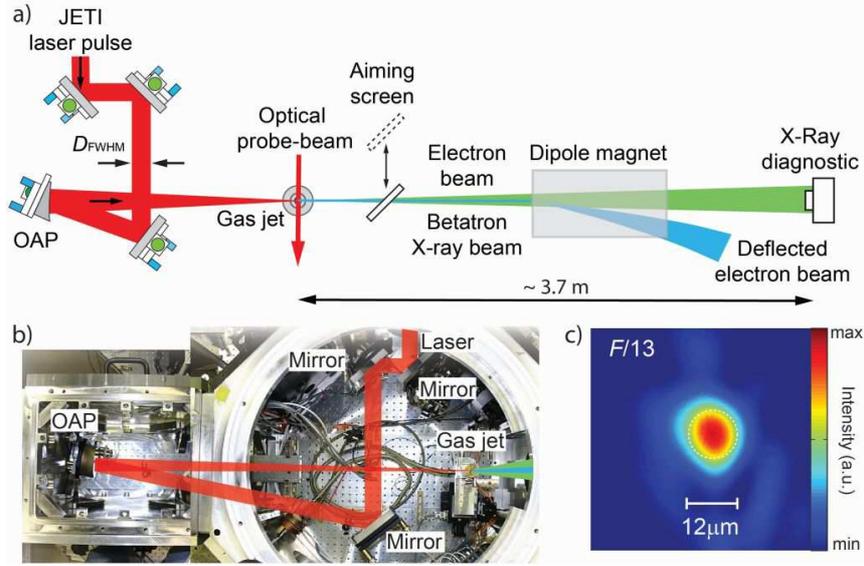

*Figure 3* (color online) Schematic of the electron acceleration and x-ray generation setup. The JETI laser pulses (red) are focused by an off-axis parabolic (OAP) mirror into a millimeter-scale, super-sonic gas jet, which leads to a maximum intensity of $7.4 \times 10^{18}$ W/cm² at FWHM. The electrons (blue) are accelerated mainly in the blow-out regime and detected with a scintillating screen and an electron spectrometer comprising of a permanent dipole magnet as an energy dispersive element. The generated x-rays (green) are recorded with an x-ray CCD detector using various types of diagnostics as described in the article. a) Schematic of the whole experiment and b) photograph of the interaction chamber (top-view). c) Focal spot for the F/13 focus optic. The dashed circle corresponds to the FWHM area.

The size of the focal area and the encircled energy is one of the most important parameters for the experiments. Our collimated laser pulse can be focused down to a focal spot diameter of 12µm (Fig. 3 c)) resulting in an intensity averaged over the FWHM area of the focus of $7.4 \times 10^{18}$ W/cm². The gas target consisted a rectangular shaped nozzle with an opening of (1.0x2.4)mm² mounted on pulsed valve fixed on an x, y and z translation stage. The nozzle could also be rotated which is necessary to align the rectangular geometry relative to the laser beam axis. With this setup we obtained a plasma with a nearly top-hat-like electron density distribution with an adjustable peak density[33] up to $(3.0 \pm 0.3) \times 10^{19}$ cm$^{-3}$. The electron beam was visualized with scintillating screens. The electrons were deflected out of the beam path by a magnetic dipole spectrometer (0.7T over 20cm). Depending on their energy they hit two different screens for detecting electrons in the range of 10-55 and 60-350MeV, respectively (Fig. 3 (a)). Along with the relativistic electrons a beam of betatron x-ray radiation was observed co-propagating along the laser-axis. For a precise characterization of the laser produced x-ray radiation a thermo-electrically cooled, back-illuminated, deep-depletion x-ray CCD camera (DO936N-BR-DD-9IN) was positioned behind the electron spectrometer (Fig. 3 (a)).

## *4. Characterization of the laser generated betatron radiation*

The x-ray CCD camera with a detector size of about (27x27)mm² is located at a distance of 3.7m measured from the gas target, collecting photons in a solid angle of $\Omega > 4.2 \times 10^{-5}$sr centred around the propagation axis. Figure 4 (a) shows the lowest measured divergence single-shot, far-field distribution of the on-axis betatron radiation. For this shot we observe a nearly rotationally symmetric Gaussian distribution with a divergence of roughly 6mrad in both transverse directions.

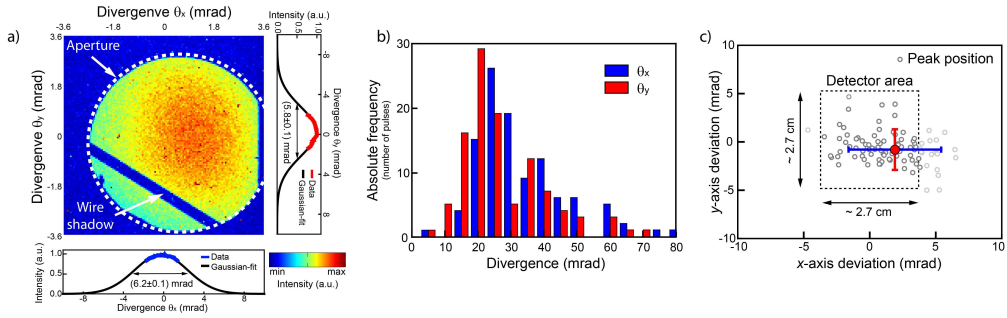

*Figure 4 (color online) a) Raw image of the angular distribution of the betatron radiation beyond 1keV recorded by the x-ray CCD 3.7m behind the laser-plasma interaction in a single-pulse. The round aperture due to the entrance of the electron spectrometer is also visible (white dashed circle) together with a shadowgram of a thin tungsten wire in the bottom left corner which indicates that the radiation originates from the x-ray source. Bottom and Right: Horizontal (blue) and vertical (red) line-outs of the beam profile fitted by a Gaussian distribution (black curve) with an emission cone on the order of 6mrad (FWHM). The electron density for this single-shot was $n_e=(1.6\pm0.2)\times10^{19}$cm$^{-3}$ for hydrogen. b) x-ray beam profile of at least a hundred individual shots indicate a clear maximum of about (25±3)mrad (blue) and (20±3)mrad (red). Here, the gas density varies around $n_e=(1.6\pm0.3)\times10^{19}$cm$^{-3}$. c) Peak position of those consecutive individual shots (grey circles) shows a shot-to-shot fluctuation of about 3.5mrad in the x-direction (blue error bar) and 2.1mrad in the y-direction (red error bar) with respect to the on-axis detector position centered at the point (0,0) (the detector area is indicated as a dashed black square). The red data point corresponds to an average over at least 100 consecutive shots, and the error bars indicate the standard error of the mean value.*

The angular distribution for about hundred individual shots is presented in Figure 4 (b). It shows a slightly asymmetric shape with an obvious maximum for a full divergence of (25±3)mrad. For several consecutive laser shots, the deviation of the x-ray beam from the laser-axis (x-ray pointing) is presented in Figure 4 (c) (grey circles). The averaged beam position of the peak (red point) was located on the x-ray CCD at an x-axis deviation of (1.9±3.5)mrad and a y-axis deviation of (-0.8±2.1)mrad. From a statistical point of view the x-ray beam profile was stable and covers the detector (grey dashed rectangle) with a large likelihood of 90%. The exact shape of the x-ray beam profile also depends strongly on the oscillation dynamics of the electrons within the accelerating plasma structure and can slightly change from shot-to-shot[16].

For a precise intensity measurement of the betatron radiation it is necessary to subtract the background noise from various sources, such as electrons hitting the nozzle or chamber walls. By summing up all background corrected events detected in a single-shot within a detection area of 1mm² (corresponds to 75x75 pixel) and taking into account the efficiency of the x-ray detector one can estimate the number of emitted photons with an energy above 1keV for different electron densities (meaning different gas backing pressures).

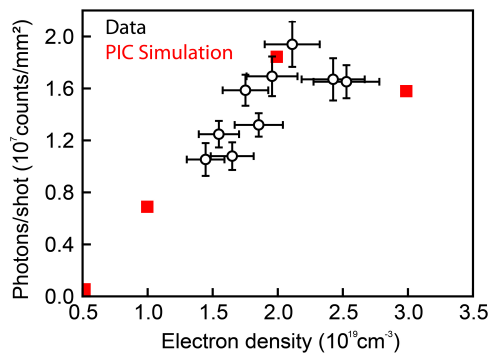

*Figure 5* (color online) *Radiated single-shot x-ray intensity beyond 1keV per area as a function of the plasma's electron density (black circles). Each data point corresponds to an average value over twenty shots. The error bars indicate the standard error of the mean value. Results from the 3D-PIC simulations (red squares) fits very well to the experimental data. Parameters used in the simulation: $n_e$=(0.5...3.0)x10$^{19}$cm$^{-3}$, λ=800nm, laser spot diameter 12μm, laser´s pulse duration 30fs and initially $a_0$=2.0 (experimental case).*

Figure 5 indicates that the x-ray intensity increases with the electron density up to a maximum. The signal then slightly decreases with increasing electron density. The maximum x-ray output was reached with a gas density of roughly $n_e=2\times10^{19}cm^{-3}$. Results from the 3D-PIC simulations, as indicated by the red squares in Figure 5, fit very well to the experimentally measured data.

So far, measurements of the betatron energy distribution were done either based on the x-ray transmission through an array of filters, the so-called "Ross filter technique"[34,35] or by using the single-photon counting method[36,37]. Progress in the investigation of both fundamental physics of secondary table top sized ultrashort hard x-ray generation and their application as non-invasive diagnostic requires powerful spectral measurement techniques to understand their generation process. Therefore it is essential to measure all important parameters of the electron and the x-ray beam simultaneously in a single-shot operation. In this article we present a careful measurement of the betatron x-ray energy distribution which provides us with both, a high energy resolution and the full intensity information of the emitted x-ray photons. This is reached by using a slit-grating spectrometer based on a transmission grating optimized to diffract x-rays from 1keV up to 20keV without attenuating the radiation by filters. Furthermore, it is shown that the experimentally determined spectra are in excellent agreement with 3D-PIC simulations.

*To give an overview of different measurement techniques we start with the Ross-filter measurement:* A single Ross-filter pair consists of two different filters materials with similar transmission curves apart from a small energy range between their K-edges. Subtracting the corresponding transmission curves results in a typically broad measuring energy range. Figure 6 (a) presents a CCD image of the x-ray radiation transmitted through various filter materials. The individual transmission curves of each filter are used to reconstruct the x-ray spectrum as shown in Figure 6 (b). The filter pairs indicated three data points for detecting photon energies in the range of (1.2±0.3)keV, (4±1)keV and (7.7±0.7)keV (grey shaded energy bands in Fig. 6 (c)).

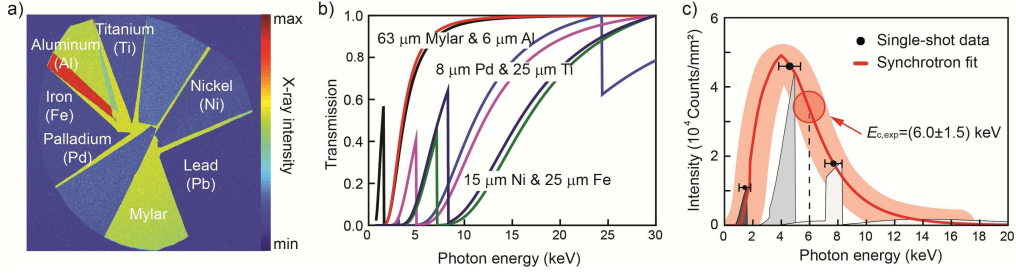

*Figure 6 (color online) Ross-filter pack used in the experiment to reconstruct the betatron x-ray spectrum. a) CCD data shows the different filter areas. b) Transmission curves of each filter. c) Experimental single-shot data (black points, error bars corresponds to the measured energy uncertainty) together with the transmission of the Ross-filter pairs (grey shaded areas). Dark grey: 6µm Al subtracted from 63µm Mylar. Grey: 25µm Ti subtracted from 8µm Pd. Light grey: 15µm Ni subtracted from 25µm Fe. The synchrotron fit (red line) with a critical energy experimentally measured at $E_{crit}$~6keV fits very well including a standard deviation (shaded red circle) of $\Delta E_{c,exp}=\pm1.5keV$ (experimental case for the single-shot data).*

There is no need to subtract the background signal when calculating the difference in x-ray signals from a Ross-filter pair, as both filters measure the same background signal which automatically cancels. The experimentally obtained single-shot data (black points) within this corresponding energy band fits very well to the theoretically predicted synchrotron radiation's spectrum (red line in Figure 6 (c)) assuming a standard deviation of $\Delta E_{c,exp}=\pm1.5keV$ (red shaded circle). The red shaded area illustrates the error bars over the measured energy range, including the error in the determined critical energy. The theoretical spectrum was calculated according to Equation 3 and the resulting curve was multiplied with the respective quantum efficiency of the x-ray CCD. Parameters used for the theoretical synchrotron spectrum (red line) were $n_e=1.6\times10^{19}cm^{-3}$, $\gamma=240$ (electron energy of 120MeV, both measured) and electron oscillation amplitude of $r_\beta=0.7µm$ (best fit parameter). A detailed description of the corresponding indirect measurement[28] for $r_\beta$ is given in the next section. The Ross-filter technique relies on the assumption that the x-ray beam has a uniform intensity distribution across the respective filters and that the energy distribution is synchrotron-like. On the one hand this method is very simple to apply and for different filters with different thicknesses[38] one can measure way above 20keV but on the other hand since the small number and energy width of the sample points are determined by the amount and transmission curves of the Ross-filter pairs (here only three, meaning 6 filters) this measurement cannot give details of the spectral shape or any deviation from a synchrotron spectrum.

A more precise measurement of the x-ray energy distribution is possible with the *Single-photon counting (SPC)* method. It provides a higher energy resolution of about 200eV over a broad energy range from 1keV to 20keV and can be easily measured in a single laser shot with an x-ray CCD camera. In the SPC mode, the number of generated electrons (the charge within the silicon CCD chip) is related to the incoming photon energy. The x-ray spectrum of the detected photons can be recovered by computing the intensity histogram of the CCD signal. Typically, the generated charge cloud is not deposited in only one single-pixel on the CCD chip (called "single-pixel event"), but rather spreads over several neighbouring pixels which is called "multi-pixel event". Several spectral reconstruction algorithms can be used which are all based on the sum over the charge recorded on all neighbouring pixels of a single-photon event. However, for the SPC method it is important to ensure that every detected photon is spatially separated from other detected photons on the CCD chip, so that only single photon events of electron-hole pair generation are recorded. In other words, this method relies on the assumption of a low photon flux regime which naturally cannot give information about the absolute x-ray intensity. The most efficient reconstruction method is based on a cluster algorithm considering all possible electron spread patterns. The residual laser-light and the low energy x-ray radiation below 1keV were blocked by a 50μm Beryllium filter in front of the CCD. Furthermore, a 0.4mm thick Mylar filter in front of the CCD was added to ensure that on average, less than one x-ray photon hit a single pixel. This low flux situation is achieved also by increasing the distance between the x-ray source and the detector rather than by attenuating the radiation using filters.

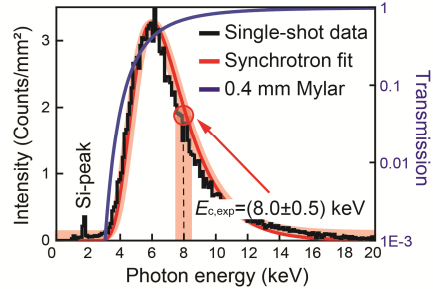

*Figure 7* (color online) Single-shot x-ray betatron spectrum obtained from the SPC method (black) including the Mylar transmission curve (dark blue). Also shown in red is the best-fit to a synchrotron distribution with a measured critical energy of (8.0±0.5)keV including a standard deviation (shaded red) of $\Delta E_{c,exp}=\pm0.5$keV. Parameters used for the synchrotron fit (red line): $n_e=1.2 \times 10^{19}$cm$^{-3}$, electron energy of 100MeV and electron oscillation amplitude $r_\beta=0.7$µm (experimental case for the single-shot data).

A typical single-shot x-ray spectrum obtained with the evaluation techniques of the SPC method is shown in Figure 7. Here, the final evaluated single-shot intensity spectrum (black) including the Mylar filter transmission curve (dark blue, right axis) and the CCD's quantum efficiency curve is shown. A 0.4mm thick Mylar filter acts as a highpass filter determining the minimum detected energy. On the high energy side of the spectrum the decreasing quantum efficiency of the CCD chip defines the detection limit. For given experimental parameters $n_e$, $\gamma$, $r_\beta$, the betatron spectrum emitted by a single electron can be described by Equation 3. As already mentioned in the Ross-filter section the theoretically predicted synchrotron radiation spectrum takes into account these filter and quantum efficiency effects and shows the adjusted data (red line) which fit very well. The red shaded area illustrates the precision of measured critical energy determination by showing the synchrotron distribution corresponding to $E_{c,exp}=(8.0\pm0.5)$keV. The measured spectrum decreased exponentially in a range from (6.0±0.5)keV to 20keV. The best-fit of the experimental measurements by a theoretical synchrotron distribution was obtained for an electron density of $n_e=1.2 \times 10^{19}$ cm$^{-3}$, $\gamma=200$ (electron energy of 100MeV) and an electron oscillation amplitude of $r_\beta=0.7$µm (red line). The experimental measurement of the transverse amplitude, $r_\beta$, of the electron within the plasma is described below in more detail.

So far, spectral measurements of the betatron radiation were described based on either the Ross-filter technique, which suffers from a lack of sample points, or by using the single-photon counting method, which requires the low photon flux regime. In this part, the energy distribution of the betatron radiation measured by a *slit-grating spectrometer* is discussed (Fig. 8 (a-d)). The basic experimental setup of the grating itself and also the spectrometer is shown in Figure 8.

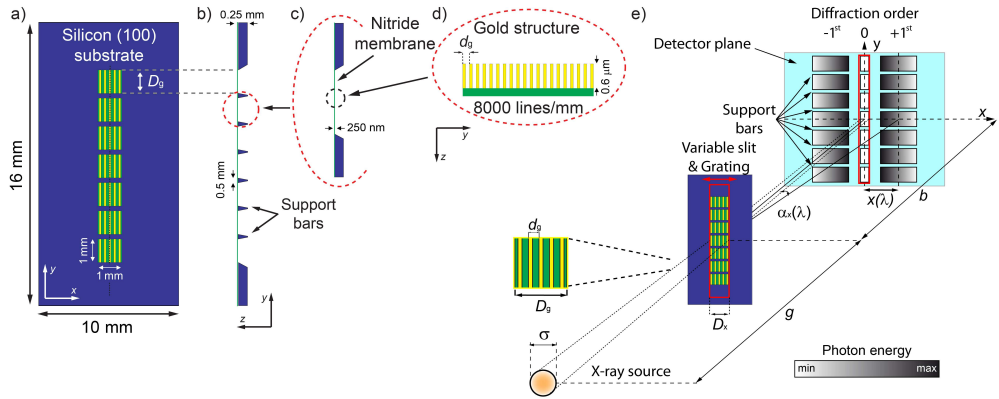

*Figure 8 (color online) Layout of the spectral measurement based on a slit-grating spectrometer. a) The whole transmission grating structure with a dimension of (10x16)mm² on a 0.25mm silicon substrate (blue). b) Cross-section indicates the 0.5mm thick support bars. c) Between the support bars is the 250nm nitride membrane (green). d) The gold bars (yellow) of 0.6µm height are placed on the nitride membrane with a period of 8000lines/mm. e) Within the detector plane (cyan blue), a schematic diffraction pattern is shown indicating the ±1$^{st}$ diffraction orders together with the projected shadow of the rectangular slit (red). The position $x(\lambda)$ of the ±1$^{st}$ diffraction orders depends on the diffracted wavelength.*

The slit-grating spectrometer was located at a distance, g, of approximately 2.3m downstream from the x-ray source. The x-ray source projects a shadow of the rectangular slit (red) onto the detector plane (cyan blue). The slit aperture was chosen in order to provide a high x-ray flux at moderate spectral resolution for single-shot measurements (wide slit) or for measurements in the multi-shot regime including high spectral resolution (narrow slit). The transmitted radiation was diffracted into the first and minus first diffraction orders under a wavelength dependent angle of $\pm\alpha_x(\lambda)$ with respect to the laser-axis. The deviation of the first diffraction order (shadow of the slit) is given by the grating formula: $\sin(\alpha_x) = \frac{\lambda}{d_g}$, where $d_g$ is the horizontal grating period and $\lambda$ the respective wavelength. For the experiment, a grating was used with $d_g$=125nm and 8000lines/mm. The location x($\lambda$) of the diffracted signal onto the detector plane can be calculated according to: $x(\lambda) \approx \lambda \frac{b}{d_g}$, using a distance from the grating to the detector plane of b=1.42m. Typical experimental values for the slit-width were $D_x$=(200…500)µm which resulted in a spectral resolution limited geometrically of about $\Delta\lambda$~0.07nm. To protect the very thin grating structure, the residual laser-light was suppressed by a 5µm diamond-like carbon (DLC) foil in front of the grating spectrograph. Real-time recording was provided by an x-ray CCD detector.

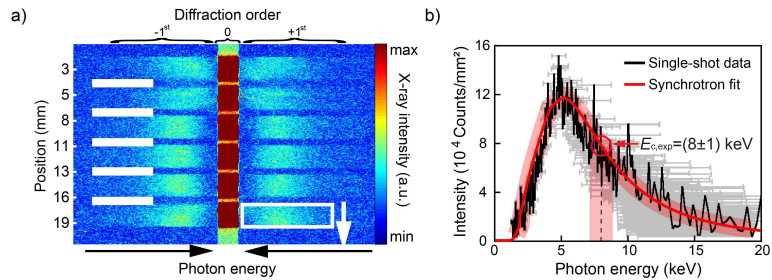

*Figure 9 (color online) a) CCD raw image of the spectral measurement averaged over 100 consecutive laser shots based on a slit-grating spectrometer including the calibrated energy axis in the horizontal direction and the position axis in the vertical direction. Vertical binning along the white arrow of only one diffracted signal within the white rectangle resulted in a single-shot radiation's spectrum. The horizontally oriented support bars are indicated as solid, thick white lines. b) Single-shot spectra of the betatron radiation measured by means of a slit-grating spectrometer (black line) and the best-fit to a synchrotron distribution (red line) with a measured critical energy of (8±1)keV including a standard deviation (shaded red) of ±1keV. The horizontal grey lines indicate the limiting spectral resolution of the slit-grating spectrometer as a function of the respective x-ray energy. Parameters used for the synchrotron fit: $n_e$=2.5x10$^{19}$cm$^{-3}$, averaged electron energy of 150MeV and electron oscillation amplitude $r_\beta$=0.7µm (experimental case for the single-shot data).*

Figure 9 (a) shows a typical raw image on the x-ray CCD detector averaged over 100 consecutive laser shots. The ±1$^{st}$ diffraction orders are clearly visible together with the grating support bars (indicating as horizontal white lines). The intensity coloured by dark blue can be attributed to the background that must be subtracted from the x-ray signal. Vertical binning of only one diffracted signal (indicated by a white rectangle) and arrangement of the energy axis in ascending order lead to a typical single-shot betatron radiation's spectrum shown in Figure 9 (b). The estimated spectral resolution (grey area) was a strong function of the respective photon energy and corresponded to a slit opening of 300µm. For the spectra, the best fit with a theoretical synchrotron distribution (red line) was obtained for a measured critical energy of about 8keV. The parameters used for the simulation were given by $n_e=2.5 \times 10^{19} cm^{-3}$, electron energy of 150MeV and electron oscillation amplitude $r_\beta=0.7\mu m$. The theoretical synchrotron fit shown in Figure 9 (b) corresponds to a "real" critical energy (after correcting for the filter functions and the CCD quantum efficiency) of 8.5keV which is in agreement to the critical energy calculated from Equation 4 for the parameters used. Note, that this measurement method is also able to highly-resolve the angular dependence of the betatron spectra in both transverse directions in a single laser shot[39].

*Comparison of the techniques for spectral measurement:*

To ensure a realistic and accurate comparison of all measured betatron spectra, we corrected for the spectral response of the filter used in the SPC spectra. Because the 0.4mm thick Mylar filter had a low energy transmission that approaches zero for low photon energies the calculated spectrum went to infinity and was only meaningful above roughly 5keV. Figure 10 shows the shape of the single-shot betatron x-ray energy distribution which was characterized by using the Ross-filter technique (yellow points), the single-photon counting method (blue curve) as well as a slit-grating spectrometer (black curve). The measured betatron spectrum was confirmed for all measurement techniques, described in the previous section. The experimental findings were well described by 3D-PIC simulations (green points) and by the formulas derived for synchrotron radiation according to Equation 3 (red curve).

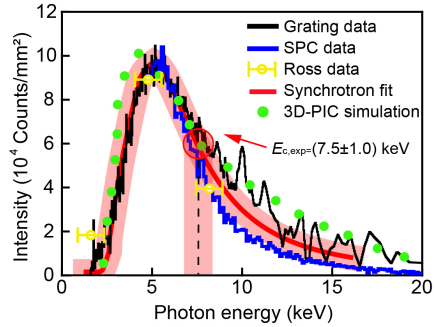

*Figure 10* (color online) Comparison of the single-shot betatron spectra measured with the Ross-filter method (yellow), the SPC technique (blue) as well as the slit-grating spectrometer (black). For a better comparability, the SPC spectrum measured within the low photon flux regime is multiplied by a factor of $\times 10^4$. The red line corresponds to the best-fit synchrotron distribution according to Equation 3. Parameters used for the synchrotron fit (red line): $n_e=2.5\times10^{19}cm^{-3}$, electron energy of 150MeV and electron oscillation amplitude $r_\beta=0.5\mu m$. The green circles correspond to a 3D-PIC simulation. Parameters used in the PIC simulation: $n_e=2.5\times10^{19}cm^{-3}$, $\lambda_L=800nm$, laser's pulse duration 30fs and an initial $a_0=2.0$.

For the comparison, the theoretical spectrum was multiplied with the respective quantum efficiency of the x-ray CCD, which was the same for all measurements. The synchrotron fit can be understood as the spectrum of the betatron x-ray source averaged over the x-ray's emission angles and the electron parameters. While the well-known Ross-filter technique (yellow points) get a rough estimation of the spectral shape, the SPC technique (blue curve) together with the slit-grating spectrograph (black curve) overcome the limitation of a low energy resolution due to a small number of sample points. The much higher energy resolution is only limited by the spectral and spatial resolution of the CCD camera or the grating efficiency, depending on the method used. Besides the advantages of the SPC method, the main drawback is the low photon flux requirement. This can be overcome by the slit-grating spectrograph whose energy range is only limited in the high-energy range, by the efficiency of the grating material as well as the quantum efficiency of the CCD camera.

To go further in the characterization of the betatron source the next step is to measure the x-ray spot size. The very first step to get a rough estimate is to back-light an object with the betatron x-rays and look at the shadow formed on the detector. If the shadow of the object is clearly visible on the detector, then the radiation originates from the laser-plasma interaction alone and the x-ray source's size cannot be larger than the object's diameter. This first estimation of the source's size is refined by comparing the recorded shadowgram with results obtained with Fresnel diffraction[17,40].

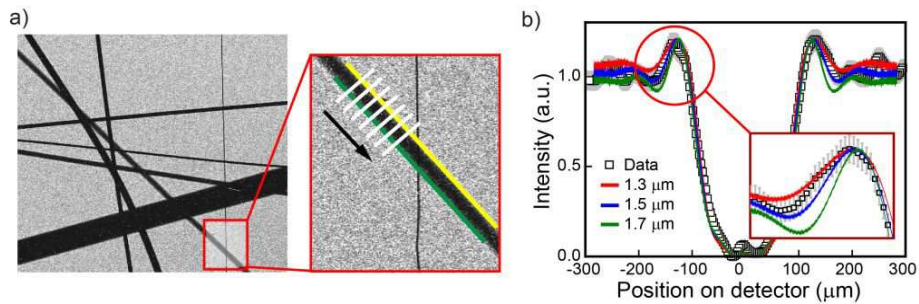

*Figure 11 (color online) Experimentally measured raw data including the intensity distribution used to measure the x-ray source's size. a) CCD raw image of the tungsten wires' shadows with different wire thicknesses ranging from 7.5μm to 100μm. The red rectangle represents the magnified region where the wire with diameter of (7.5±0.1)μm is still clearly visible. Also shown are the computed upper (yellow) and the lower (green) border of the wire's edge and the computed line-outs (white) from which one can evaluate the intensity distribution. b) Measured (black squares) and simulated (colored lines) intensity distribution on the detector using Fresnel diffraction from a radiation source with a measured broad-band betatron spectrum including a measured critical energy and a Gaussian intensity distribution. The best-fit for the amplitude and the width of the first fringe is given by a source size of σ=1.5μm (blue curve). The grey error bars represents the error band of ±0.3μm.*

For the experiment described in this article, the Fresnel number was calculated to be much larger than one which validates the near-field description of the x-ray intensity distribution at the detector. Starting at the source plane, the electric field was calculated at the position of the wire which was described by its cylindrical shape, complex index of refraction, and a negligible surface roughness. Accordingly, one can calculate the intensity distribution at the detector plane from the electric field modified by the wire. Since Fresnel theory is monochromatic, the intensity distribution at the detector was simulated for different photon energies while taking into account the energy dependent refractive index of the wire. The monochromatic images were summed together with a weighting that followed the shape of the measured betatron spectrum. According to Fresnel diffraction theory, the source's intensity distribution is fully determined by the setup, i.e., the object and image distances, the shape of the wire, and the x-ray spectrum. It follows that the accuracy of the retrieved source size is mainly limited by measurement errors of these quantities. Figure 11 (b) shows an averaged line-out of the measured intensity distribution (black squares) and the best-fit for the first overshot (blue curve, close-up image). For a broad betatron spectrum, only the first overshot is clearly visible and the smaller the Gaussian-like source size, the higher the first overshot. Thus, the source's size can be calculated using the first overshot at the edge of the wire's shadow. Assuming a Gaussian intensity distribution of the x-ray source, which was justified by the Gaussian-like beam profile in the far-field (Fig. 4 (a)) the Fresnel diffraction modelling revealed an upper limit for the radiation source size of only $(1.5\pm0.3)$µm FWHM. This small spot size of the betatron radiation is very encouraging for applications requiring an x-ray source with high peak-brightness.

*Single shot polarization measurement:* In the following part, a well-designed x-ray polarimeter with matched characteristics is described. The general experimental layout was shown in Figure 3 (a). To measure the betatron radiation's polarization state in a single-shot operation, this setup was extended by placing the polarimeter arrangement between the electron spectrometer and the on-axis x-ray detector, shown in Figure 12.

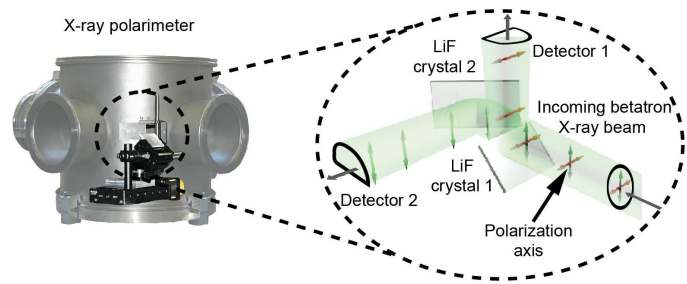

*Figure 12* (color online) Details of the single-shot x-ray polarimeter (side-view). Left: The whole polarimeter was placed inside a vacuum chamber to prevent x-ray absorption in air. Right: Schematic of the crystal orientation shows crystal 1 is irradiated only by the bottom half of the betatron x-ray beam while crystal 2 reflects only the upper part. Also shown, the polarization axis of the incoming betatron x-ray beam (crossed double arrows). Crystal 1 only reflects the horizontally polarized photons (red double arrows) and crystal 2 only the vertically polarized photons (black double arrows).

The x-ray polarimeter consists of two Lithium-fluoride (LiF) crystals using the strongest Bragg reflection 200. For ideal mosaic crystals, the diffracted polarization components with the electric field components parallel (π-component) and perpendicular (σ-component) to the diffraction plane are defined by the incident and diffracted wave vector amount $\cos^2(2\theta)$ and 1, respectively[41]. The 200 netplane distance ($2d_{200}$=4.027 Å) fits perfectly to the maximum of the betatron spectrum by using the Bragg condition ($\lambda = 2d_{hkl} \cdot \sin\theta$, where λ is the reflected wavelength and $d_{hkl}$ the lattice distance of the crystal reflection with hkl - diffraction indices) close to the Bragg angle of 45°. At this angle only linearly σ-polarized x-rays are diffracted. In order to increase the diffraction efficiency, the crystals' perfection was reduced by a specially controlled grinding procedure. With this treatment the crystals provide integrated reflectivities which are more than nine times higher than for a perfect crystal. The reflectivity was determined to be constant within 6% over the crystal's reflecting area. The FWHM of the experimentally determined rocking curves yield an energy resolution of (13…22)eV at 4.6keV photon energy. Both crystals were built into the polarimeter so that their dispersion planes were mutually perpendicular, allowing detection of the vertical and horizontal polarization states of the betatron radiation with an extinction ratio of 1:33 (3.5%). This assumes an alignment accuracy better than 1° for the Bragg angle and uses the measured divergence of the radiation and width of the crystals' reflection curves. After measuring the beam's pointing characteristics (Fig. 4 (b)), x-ray polarizers were added to the experimental setup. Crystal 1, which reflected the horizontally polarized photons (red double arrows, detected by detector 1) was irradiated by only one-half of the x-ray beam while the other part was reflected by crystal 2 which reflected the vertically polarized photons (black double arrows, detected by detector 2). That means detector 1 only detected horizontally polarized x-ray radiation and detector 2 only detected vertically polarized x-ray radiation.

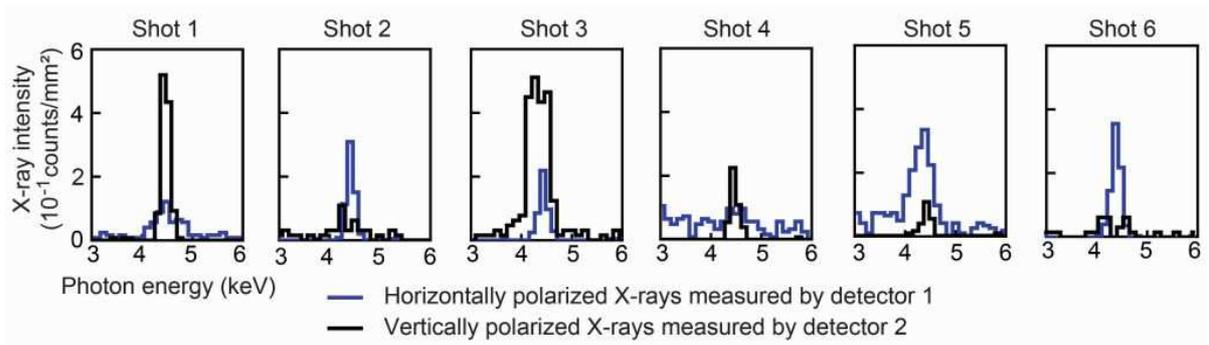

*Figure 13* (color online) *Typical single-shot Bragg reflection measurements. The betatron x-ray radiation for different laser shots differed significantly in terms of their direction of polarization (absolute value for the blue and black curve). They are vertically polarized (black line, detected by detector 2) or horizontally polarized (blue line, detected by detector 1). The photon number within only one Bragg-peak corresponds to nearly $10^3$ photons/shot. The shot-to-shot fluctuations contribute to the asymmetric intensity distribution inside the focal spot which vary from shot-to-shot as well.*

Figure 13 shows the experimentally obtained crystal reflection from detector 1 (blue line) and detector 2 (black line), respectively[30]. The reflected peak was located at 4.6keV corresponding to the Bragg-condition. For different laser shots, the measured x-ray spectra differed significantly in terms of the number of photons which were reflected. Or in other words, the polarization direction of the betatron radiation changed from shot-to-shot. Furthermore, one can conclude that in the bubble acceleration regime[10,21,42] the polarization state of the betatron radiation is independent of the laser's polarization direction which was not changed during the whole experiment. The driving laser pulse was polarized in the horizontal plane and the direction of polarization of the generated betatron radiation changed from horizontal to vertical on a shot-to-shot basis. The observed characteristics of the betatron radiation's polarization were also confirmed in 3D-PIC simulations which took into account the electron dynamics within the plasma wave. Hence, the theoretical findings predicted by the PIC simulation will be discussed later. This polarization experiment showed for the first time, that betatron x-ray beams have defined polarization features and the detailed results will lead to a deeper understanding of x-ray generation within a laser wakefield accelerator. Combining this knowledge with ultrashort, broad-band keV x-rays from a compact laser-plasma source will help pave the way for an abundance of applications.

**5. Application of betatron radiation for non-invasive measurement of the injection and acceleration mechanism in laser-plasma accelerators**

In this final section we report on using betatron radiation for the first time to gain insight into the electron injection into the plasma waves and the subsequent electron acceleration dynamics that so far have not been studied in detail[28,30,40,43]. For this type of investigations, it is essential to measure all important parameters of the electron and the x-ray beam simultaneously in a single-shot operation. Parameters to be measured include the plasma density, the electron energy, the x-ray energy distribution, and the x-ray polarization state. In the following we describe how knowledge of the above mentioned (and also measured) parameters allows to estimate other important properties of the betatron radiation including the transverse oscillation amplitude, $r_\beta$ of the betatronic electron motion and reveal the role of off-axis electron injection for the subsequent electron trajectory.

*Amplitude of the electron oscillation inside the plasma wave:* During the laser-plasma interaction, the electrons undergo oscillations and emitting betatron radiation[23,44] (Fig. 1). The x-ray photons are mainly emitted from the region in the plasma where the electrons have the maximum energy[12,28] and therefore the measured parameters, in particular the oscillation amplitude can be understood as weighted averaged, i.e. the signal is emitted mainly from the last few oscillations. Using a free-expanding gas jet and setting the electron injection position correctly, the highest electron energy is accompanied by a narrow electron spectrum (Fig. 14 (a)). To get quasi monoenergetic electron spectra for different energies the presented measurements were performed for slightly different electron densities of $n_e=(1.0…2.5) \times 10^{19} cm^{-3}$ and an optimized laser focal spot positioned at the beginning of the steep electron density profile. Optimized conditions in this context mean that the backing pressure and the position of the laser focus with respect to the gas density profile were chosen correctly[45]. Otherwise, for a slightly different plasma density and an un-optimized target position, meaning the laser's focus position relative to the gas's density distribution, the electron and x-ray beam was much weaker and exhibited a much larger divergence. For the evaluation of the experimental data only x-ray spectra generated from quasi- mono-energetic electron beams were selected, as shown in Figure 14 (a). Based on experimental findings[7] and results obtained from PIC simulations carried out in the context of this work, self-injection must take place at a position where the electrons in the remaining plasma length are only accelerated and no dephasing takes place[62]. If the injection occurs too early, much broader electron spectra with reduced peak energy due to dephasing are expected. The peak energies in Figure 14 (a) ranged from 65MeV (#1, #2) to 115MeV (#6) with a FWHM between 1MeV and 3MeV. Taking into account the respective filter functions, the CCD quantum efficiency and the measured background signal one can calculate the betatron spectrum from the measured single-event spectrum as shown in Figure 14 (b).

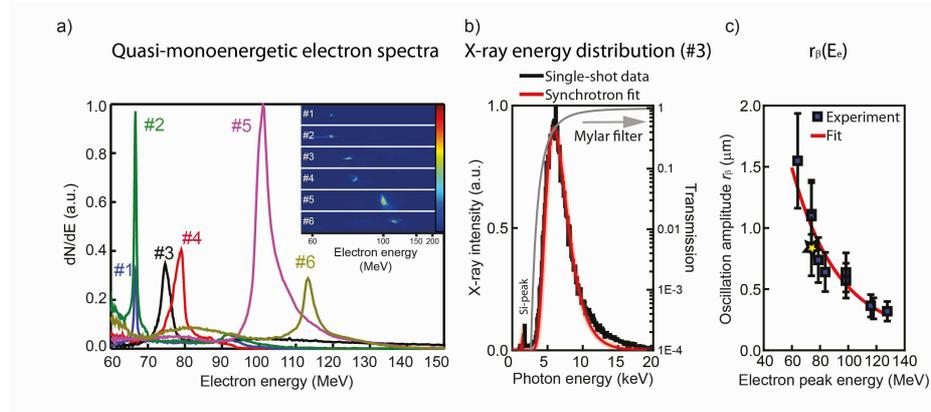

*Figure 14* (color online) a) Measurement of quasi-monoenergetic electron spectra for different electron densities $n_e=(1.0...2.5)x10^{19}cm^{-3}$ showing narrow spectral features with an energy ranging from (65...115)MeV. Inset: Raw data from the scintillator screen of the electron spectrometer. b) Normalized single-shot betatron spectra (black curve) corresponding to the electron spectra #3 of part (a) in a range limited by the transmission of a 0.35mm thick Mylar filter for low photon energies (grey line) and the reduced detection efficiency of the CCD for high photon energies. The red curve illustrates the precision over the measured critical energy determination by showing the theoretically predicted synchrotron distribution according to Equation 3 (multiplied by the filter's transmission curve and the CCDs response curve). Parameters for the simulations were $n_e=1.2x10^{19}cm^{-3}$, electron peak energy of $E=75MeV$ (experimental case) and a betatron oscillation amplitude of $r_\beta=(0.9\pm0.3)\mu m$. c) Deduced betatron oscillation amplitude $r_\beta$ as a function of the electron's peak energy. The red curve is to guide the eye. The highlighted point (yellow star) indicates the data set #3 of part (a) and (b) mentioned above.

Here, a typical measured single-shot x-ray spectrum is shown (black curve). The spectral betatron distribution indicated an experimentally determined critical energy of $E_{c,exp}=(8.0\pm0.5)keV$. The recorded x-ray spectrum was mainly limited by the transmission of the strongest absorbing filter in the beam path. Since the plasma density, $n_e$, and the peak-electron energy were measured for each laser shot simultaneously, the theoretical x-ray spectrum according to Equation 3 (taking into account the filter's and CCD's transmission functions) could be fitted to the measurement with the oscillation amplitude, $r_\beta$, as a free fit parameter (red curve in Fig. 14 (b)). Note that for this evaluation only strongly peaked electron spectra were taken into account where the assumption of a narrow energy spread was justified. For the electron energy distribution #3 of Figure 14 (a) and the corresponding single-shot betatron spectrum shown in Figure 14 (b) (black curve) the best agreement was achieved for an oscillation amplitude of $r_\beta=(0.9\pm0.3)\mu m$ (highlighted star in Fig. 14 (c)). By varying the gas density, which affected the electrons' peak energy, the amplitude of the betatron oscillation as a function of the electron peak energy could be deduced as shown in Figure 14 (c). Each measurement point corresponds to a set of data ($n_e$, $\gamma$, $r_\beta$) for a single laser shot. A higher electron peak energy, i.e., an increased electron mass, $m_e$, results in a smaller oscillation amplitude, $r_\beta$, which reaches a minimum for the highest electron energy[28,29].

*Numerical modelling:* The experimental findings were also observed in a 3D-PIC simulation[28,31]. The results are summarized in Figure 15, where the integrated number of emitted x-ray photons (green curve), their peak-energy (blue curve) and the relative spectral width of the electrons (black curve) are shown as a function of the propagation length during the interaction between the laser pulse and the plasma.

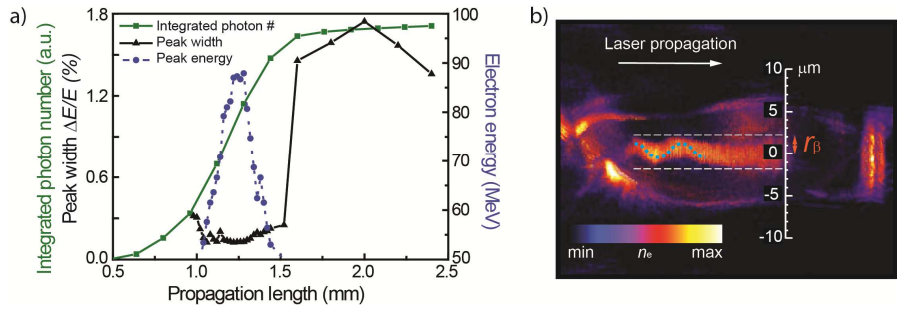

*Figure 15 (color online) 3D-PIC simulation of the amplitude of the laser-driven betatron oscillations. a) Only the fastest electrons (blue line) contribute to the main emission of the betatron signal (green line) and only the fastest electrons are peaked (black line). b) Snapshot of the electron distribution. The driving laser propagates from left to right (white arrow). The simulation confirmed the experimentally deduced value of the transverse electron oscillation amplitude of $r_\beta<1\mu m$ (orange double arrow).*

It is shown that during the interaction the energy spectrum of the injected electrons evolves from an initially broad distribution to narrow peaks (black) with increasing peak energy (blue). The simulations revealed that the largest number of x-ray photons (green) is indeed emitted within a very limited region in the plasma where the electrons' peak energy is at a maximum and the relative energy width is at a minimum. Once the electrons start to enter the dephasing region, the spectrum becomes broader again and the emission of the betatron signal drops. This means that the majority of the x-ray photons are generated over a short distance over which it is a good assumption that the electron energy stays approximately constant. Since self-injection of electrons into the plasma wave is a highly non-linear process (which was not actively controlled during the experiment) an experimentally measured narrow energy spread of only a few percent indicates that the plasma ended before the electrons could enter the dephasing region[46]. Otherwise, the electrons' energy spectrum would have been much broader. Since only those shots showing an energy spread of a few percent were specifically chosen for further analysis, these electrons obviously did not suffer from dephasing. It is claimed that in these cases the majority of the experimentally detected betatron x-ray photons was indeed emitted from only a short length within the plasma; in this case right at the end of the interaction in the plasma. Figure 15 (b) shows a snapshot of the electron density distribution. The simulations confirmed the experimentally deduced value for the micrometer-scale, transverse oscillation amplitude of the electrons inside the plasma wave.

*All-optical control of the x-ray polarization state:* Finally in this article we investigate the correlation of the already measured single-shot x-ray polarization with simultaneously measured properties of the electron bunch, mainly its energy and vertical spatial distribution. It turns out that one can classify the betatron radiation's polarization states with respect to the electron distribution into two linearly polarized, orthogonal states. Furthermore, it was demonstrated that the polarization of the betatron radiation can be controlled by spatial shaping of the driving laser pulses[30,47]. In the following we describe, both experimentally and theoretically, how the control of the electrons' trajectories, and hence control of the betatron radiation's polarization state, by using either an asymmetric intensity distribution in the laser focus, or by tilting the laser's pulse-front[48]. The data were once again measured for each individual laser shot: the polarization state of the betatron radiation, a shadowgram of the plasma wave, the electrons' energy spectrum, and the electrons' transverse beam profile in one dimension. These simultaneously recorded data allowed for the correlation of the polarization states presented in Figure 13 to the respective electron spectra. Besides the desired quasi mono-energetic single-shot electron spectra typical shapes indicate either a wave-like trace of the electrons on the scintillating screen or exhibit a narrow transverse electron distribution. Depending on the exact longitudinal and transverse position of the electron injection, the following parameters vary from shot-to-shot: the net acceleration length, the plane of the betatron oscillation, the electrons' energy spectra, the electrons' spatial distribution, and the x-ray's polarization state. Here, one benefits from the fact that the electron spectrometer disperses the electron pulse along the horizontal axis, whereas the signal is spatially resolved along the vertical axis[49]. The wave-like electron distribution can be explained by electrons oscillating mainly in the vertical direction while the laterally smooth structure indicates an additional collective wiggling of the electrons in the horizontal direction.

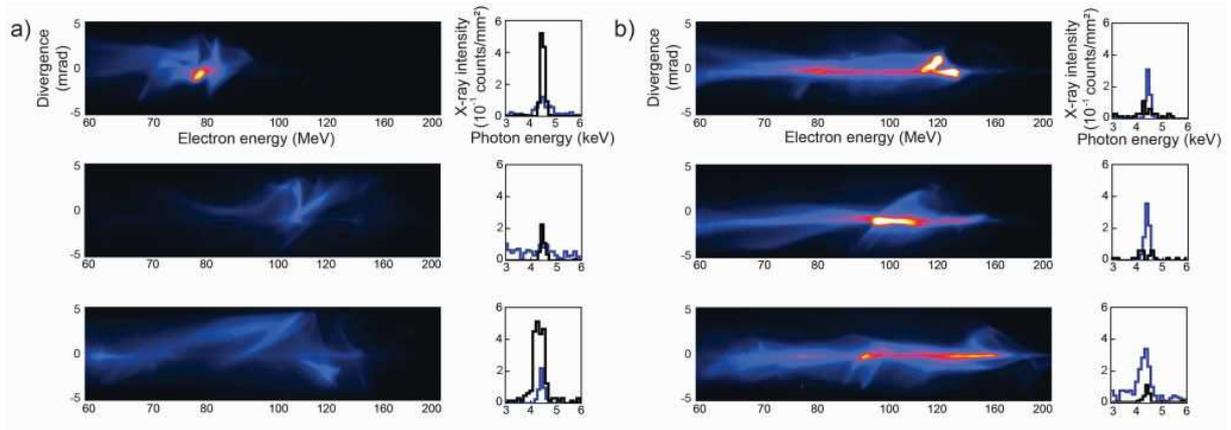

*Figure 16* (color online) *Electron spectra and simultaneously measured x-ray polarization state. Typical single-shot electron spectra on the high-energy scintillating screen exhibit either a wave-like trace (a), or are vertically well confined (b). Note, that only the high-energy electrons contributed to the generation of observable betatron radiation. For the wiggled electron traces (a), the x-ray radiation is mainly vertically polarized (black line) and has only a minor contribution in the horizontal polarization direction (blue line). For the vertically confined electron traces (b) the horizontal polarized component dominates. The shot-to-shot fluctuations are attributed to the asymmetric intensity distribution inside the focal spot which vary from shot-to-shot as well.*

Correlating the betatron polarization states to the electron spectra showed that for the wiggled electron trace the simultaneously measured x-ray signal was primarily polarized in the vertical direction (Fig. 16 (a-c)) whereas for the straight electron trace the x-ray signal was primarily polarized in the horizontal direction (Fig. 16 (d-f)). This behaviour can be explained by looking at the origin of the x-ray radiation, specifically, the oscillating movement of the relativistic electrons within the plasma wave (Fig. 1). This means that the radiation characteristics in particularly the polarization state depend critically on several electron parameters, for example the exact longitudinal and transverse position of the electron injection, the net acceleration length, and the plane of betatron oscillation. These parameters are likely to vary from shot-to-shot which influenced the direction of the x-ray polarization. Unfortunately, not all of the parameters were at the direct disposal of the experimentalist, such as the laser's pulse parameters and/or gas density fluctuations. Self-injection is a highly non-linear process and even small shot-to-shot fluctuations of the laser's parameters (e.g., energy and/or pointing) result in different electron injection parameters and different betatron polarization states which is shown in Figure 13. For the experiment described in this article, the main effect influencing the self-injection of electrons was the asymmetric intensity distribution of the focal spot, which varied from shot-to-shot (Fig. 17).

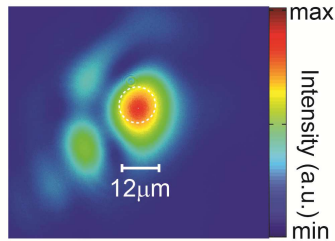

*Figure 17* (color online) *The main contributing effect of the electron off-axis injection into the plasma wave is due to an asymmetric intensity distribution inside the focal spot which generates an asymmetric wakefield.*

Since the laser intensity profile was asymmetric, the radial ponderomotive force was different in each direction, driving an asymmetric plasma wave resulting in a substantial off-axis electron injection. Since the polarization of the emitted betatron radiation is determined by the plane of oscillation of the electrons injected into the plasma wave, the measured x-ray polarization provides additional information about the injection of the electrons into the plasma wave.

3D PIC simulations for a tilted pulse-front and a non-symmetric intensity distribution of the focal spot were carried out. Both deviations from a perfect Gaussian beam were expected to inject electrons off-axis into the plasma wave and hence both effects are promising parameters for controlling the betatron radiation's polarization. To further investigate this effect, the asymmetry within the focal spot was increased by introducing a vertical pulse-front tilt which was sufficiently large to overcome the statistical fluctuations already seen in Figure 13 and enabled the all-optical steering of the off-axis electron injection into the plasma wave. Experimentally, the pulse-front can be tilted by slightly misaligning one of the compressor gratings[50,51], which are used in a CPA laser-system. The misaligned grating causes an angular chirp in the spectral domain corresponding to a tilted pulse-front in the time domain. Having the ability to measure the betatron radiation's polarization state in a single-shot is important for reducing sensitivity to shot-to-shot fluctuations. This is particularly important as the output signal depends critically on the exact position of the non-linear wave-breaking i.e., the electron injection position and inevitable intensity fluctuations within the asymmetric laser focus. In Figure 18 (a) an image of the plasma wave taken with the probe-beam[52] that crosses the plasma at an early stage is presented.

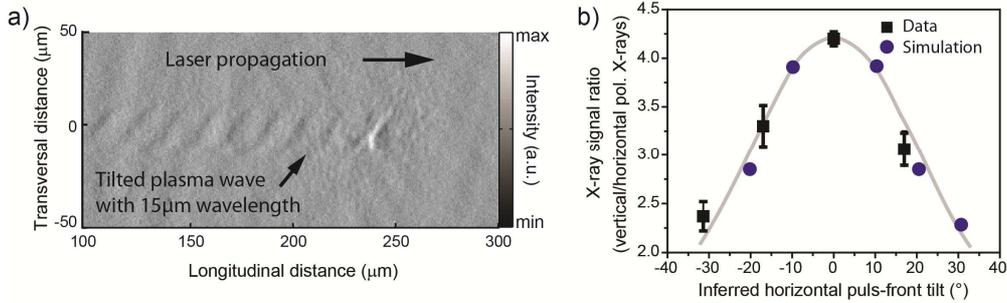

*Figure 18 (color online) Controlling the x-ray polarization by tilting the laser's pulse-front. a) Snapshot of the electrons' density gradient taken with a probe-pulse with 6fs pulse duration for zero horizontal angular dispersion. The laser pulse propagates from left to right (black arrow) and the plasma wave is vertically tilted due to a spatial-temporal asymmetry of the focal spot. b) Ratio of the vertically and horizontally polarized x-ray contribution as a function of the horizontal angular dispersion of the laser pulse. Each data point (black squares) corresponds to an average over at least 100 consecutive shots, and the error bars indicate the standard error of the mean value. The experimental findings fit very well to the predictions from the 3D-PIC simulations (blue squares). The grey line is to guide the eye.*

The gas jet is placed in front of the focal position of the main laser pulse propagating from left to right. The intensity modulation (grey scale) can be associated with the plasma wave which exhibited a very strong tilt against its propagation direction. The observed spatial intensity modulation was created from the spatially-homogeneous probe-pulse refracting through the electrons' density modulation inside the plasma[53]. From the shadowgram shown in Figure 18 (a) one can estimate the wavelength of the plasma wave which is in the order of 15μm corresponding to an electron density of $0.5 \times 10^{19} cm^{-3}$. This is in good agreement with the measured density derived from an independent characterization of the density distribution in the proximity of the nozzle with an interferometer[33]. The effect of this introduced pulse asymmetry on the polarization of the betatron emission is depicted in Figure 18 (b) obtained by averaging over 100 consecutive laser shots. To reduce the effect of fluctuations of the x-ray intensity between measurements of different pulse-front tilts, the ratio of the spectrally and temporally integrated signal in the vertical plane versus the horizontal plane was plotted. The spectral integration spans the high reflectivity range of the LiF crystal in the range of (4…5)keV. Here, for a primarily vertical asymmetry in the focal spot, a net 4:1 ratio for vertical polarized x-rays arose. The vertically tilted pulse-front, causing self-injection of electrons with a vertical offset resulted in vertically polarized x-rays. To demonstrate the influence of the tilt, one of the compressor gratings was rotated in the dispersive plane away from the position with zero angular tilt, which resulted in a horizontal angular dispersion. This favourable injection in the plane with the tilted pulse-front (horizontal plane) and this ratio started to decline in agreement with our theory. However, it was not possible to completely invert the intensity ratio because the electron acceleration became more and more inefficient due to the longer pulse duration within the focal spot. Figure 18 (b) reveals the strong impact of the pulse-front tilt on the betatron radiation's polarization state. With increasing horizontal pulse-front tilt the electrons are injected horizontally off-axis with a higher probability, which increased the amount of horizontally polarized x-rays. This measurement demonstrates the feasibility of controlling the x-ray's polarization state with an all-optical method, and is in excellent agreement with the predictions based on 3D-PIC simulations presented in the following.

*Numerical modelling:* During every simulation step the change of momentum was calculated for every electron- and ion-macro-particle. According to classical synchrotron radiation (omitting QED), the number of photons emitted, their energy, their polarization state and their propagation direction can be calculated from the change of momentum. Further it is assumed, the emitted photons do not interact with particles or fields during the simulation. Simulations were carried out for a tilted pulse-front and a non-symmetric intensity distribution of the focal spot. As already mentioned, both deviations were expected to inject electrons at an off-axis position into the plasma wave. Three different kinds of laser pulses were used: a) a pulse with a Gaussian envelope in the transverse direction and a cosine shape in longitudinal direction. b) same pulse as in a) but the longitudinal coordinate in the cosine was modified to create a tilted pulse-front. c) two pulses as described in a) overlapping in time and space, but with different intensity. Combining a main pulse with exactly the same as for a) with the weaker one (only about 1/10 of the intensity) allows us to model the elongation of the asymmetric pulse.

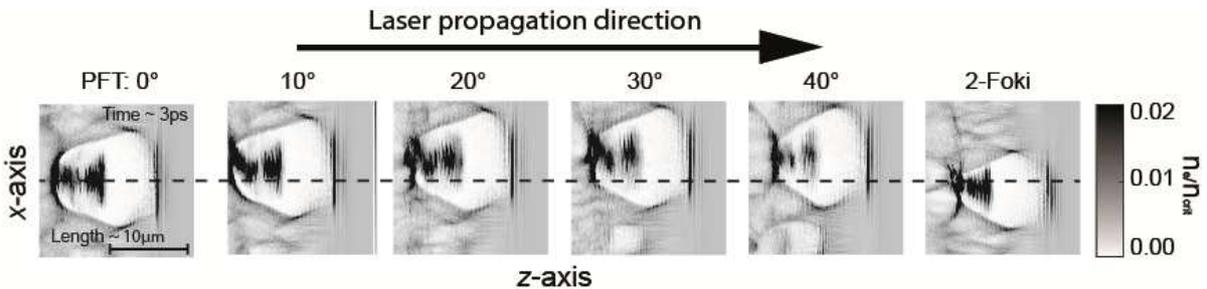

*Figure 19 Snapshots of the electrons' off-axis injection after 3ps laser propagation time simulated with the 3D-PIC code "VLPL". On-axis electron injection driven by a laser pulse with no pulse-front tilt and an ideal symmetric focal spot (PFT=0°). With a substantial pulse front tilt in the range of 10° to 40° we observe an asymmetric electron injection. The black dashed line indicates the on-axis position. An asymmetric plasma wave is also driven by a non-symmetric intensity distribution of the focal spot leading to a notable off-axis electron injection and also to collective electron-betatron oscillations (2-Foki). Simulations were carried out for the experimental parameters mentioned in the text.*

As shown by the simulations in Figure 19, without any aberration of the laser pulse the injected electrons oscillated around the laser-axis with the betatron period (Fig. 19; PFT=0°) and emitted x-ray radiation with no preferred polarization state. With an additional pulse-front tilt, for example, in the direction of the laser's polarization axis, the electrons were injected off-axis leading to a larger initial amplitude of the betatron oscillation and a significantly higher number of x-ray photons polarized in the plane of the electron oscillation (Fig. 19; PFT=10°…40°). Here, the x-ray polarization was parallel to the laser's polarization axis. Furthermore, if the pulse was tilted in the direction perpendicular with respect to the laser polarization, the x-ray and laser polarization were perpendicular to each other. These numerical results confirmed that the polarization states of the emitted betatron photons can be controlled by tilting the pulse-front of the driving laser pulse. Similar results can be achieved by introducing an asymmetric focal spot, which also drives an asymmetric plasma wave (right in Fig. 19).

## 5. Conclusion and Outlook

In this article we have provided a detailed characterization of the hard x-ray betatron radiation by measuring its radiated intensity, energy distribution, far-field, beam profile and source size. Furthermore, the betatron polarization state was found to be linearly polarized. The experimentally estimated number of $2 \times 10^7$ x-ray photons/mm² above 1keV in each shot in combination with the small spot size of (1.5±0.3)µm and the low averaged divergence of (25±3)mrad results in a peak-brightness of $10^{22}$ photons/(s mrad² mm² 0.1% bandwidth). A laser pulse duration of 30fs as an upper limit of the x-ray pulse duration was used according to simulations[54,55]. It was shown that betatron radiation can be used as a non-invasive diagnostic tool to retrieve information on the electron acceleration dynamics within the plasma wave. Measuring all the characterizing parameters of the betatron source simultaneously helped to gain information about the energy dependent oscillation amplitude of the electrons inside the plasma wave. Additionally, it is shown that the polarization state strongly depend on the location of the electron injection within the plasma wave. This leads to a detailed study of the orientation of the electron trajectory within the plasma interaction. Controlling the injection position of the electrons demonstrated the ability to tune the polarization state of the emitted x-rays. It was verified both experimentally and theoretically that the control of the electron trajectories and hence their polarization state can be realized by either an asymmetric intensity distribution in the laser's focus or by tilting the laser's pulse-front.

Such a source of hard and well polarized x-ray pulses will pave the way for studying the dynamics of magnetization via linear magnetic dichroism or for studying structural changes in thin films employing polarization dependent x-ray absorption fine structure spectroscopy. Supplemental, controlling the electron trajectory inside the plasma wave will also have a major impact on the ongoing efforts towards the realization of novel, laser-based particle accelerators. Laser-plasma accelerated electrons will be used to seed free electron lasers (FEL) which could open the way for the production of intense x-ray beams in a relatively compact system in comparison to today's intense x-ray sources. The experimental findings presented in this article, particularly the measurement and control of the electron trajectory within the plasma wave, will be of critical importance to reach the goal of feeding the electrons into an additional conventional accelerator or a permanent magnet-based undulator for generating x-ray radiation. This very new idea (FACET - Facility for Advanced Accelerator Experimental Tests)[56,57] aims to shrink the size and costs of further particle accelerators. Furthermore, the scaling of a betatron source is manifold in terms of number of photons, divergence, and spectral energy range. The route toward higher radiated x-ray energies is given by increasing the laser intensity and/or decreasing the plasma density. This can be achieved, for example, by using Petawatt-class lasers together with targets such as capillaries[8] that facilitate laser guiding and stable electron injection mechanisms such as external-optical or colliding-pulse injection[7] and density-gradient injection[58,59]. Alternative schemes to produce brighter betatron radiation and higher photon energies is given by the use of plasmas with tailored density distributions[60,61]. By controlling the plasma's density, it is possibly to control the amplitude of the electron oscillation. Moreover, the experimental findings presented in this article bring up further questions that encourage additional investigation in the fast evolving field of electron acceleration via a laser-plasma interaction.

**Acknowledgements**

This study has been sponsored by the DFG (grant TR18 A12, B9 and GRK1203), by BMBF (contracts 05K10SJ2 and 03ZIK052), by TMBWK (grants B154-09030, B 715-08008) and by the European Regional Development Fund (EFRE). Experimental support from B. Beleites, W. Ziegler and F. Ronneberger is acknowledged. M.S. appreciates fruitful discussions with Robert Lötzsch and Tino Kämpfer. We also gratefully acknowledge the contributions of Alexander Pukhov to the theory.